\documentclass[%
reprint,
superscriptaddress,
 amsmath,amssymb,asmart,
pra,
]{revtex4-2}

\usepackage{graphicx}
\usepackage{dcolumn}
\usepackage{bm}
\usepackage{xcolor}
\usepackage{bbm}
\usepackage{physics}
\usepackage{siunitx}
\newcommand{\SIadj}[2]{\SI[number-unit-product={\text{-}}]{#1}{#2}}
\newcommand{\SIpm}[3]{\SIrange[range-phrase={\text{~$\pm$~}}]{#1}{#2}{#3}}
\usepackage{multirow}
\usepackage[labelfont=sc]{caption}
\usepackage{todonotes}
\usepackage[normalem]{ulem}
\definecolor{todoxcolor}{HTML}{11aa00}

\usepackage{appendix}
\usepackage{soul}
\usepackage{epsfig}
\usepackage{float}
\usepackage{multirow}
\usepackage{braket}
\usepackage{lipsum}
\usepackage{hyperref}
\hypersetup{
    colorlinks=true,
    linkcolor=blue,
    filecolor=magenta,      
    urlcolor=blue,
    citecolor = blue,
}

\makeatletter
\newcommand{\specificthanks}[1]{\@fnsymbol{#1}}
\makeatother
\begin{document}

\title{Strong coupling at room temperature with a centimeter-scale quartz crystal}

\author{Davide Tomasella}
\altaffiliation{These authors contributed equally to this study.}
\author{Santiago Tarrago Velez}
\altaffiliation{These authors contributed equally to this study.}
\author{Sissel Bay Nielsen}
\altaffiliation{These authors contributed equally to this study.}
\author{Joost Van der Heijden}
\author{Ulrich Busk Hoff}
\author{Ulrik Lund Andersen}
\email[Corresponsing author: ]{ulrik.andersen@fysik.dtu.dk}
\affiliation{Center for Macroscopic Quantum States (bigQ), Department of Physics, Technical University of Denmark}
\date{\today}

\begin{abstract}
Brillouin-based optomechanical systems with high-frequency acoustic modes provide a promising platform for implementing quantum-information processing and wavelength conversion applications, and for probing macroscopic quantum effects. Achieving strong coupling through electrostrictive Brillouin interaction is essential for coupling the massive mechanical mode to an optical field, thereby controlling and characterizing the mechanical state. However, achieving strong coupling at room temperature has proven challenging due to fast mechanical decay rates, which increase the pumping power required to surpass the coupling threshold.
Here, we report an optomechanical system with independent control over pumping power and frequency detuning to achieve and characterize the strong-coupling regime of a bulk acoustic-wave resonator. 
Through spectral analysis of the cavity reflectivity, we identify clear signatures of strong coupling, i.e., normal-mode splitting and an avoided crossing in the detuned spectra, while estimating the mechanical linewidth \( \Gamma_m/2\pi~=~\SI{7.13}{MHz}\) and the single-photon coupling rate \( g_0/2\pi~=~\SI{7.76}{Hz}\) of our system. 
Our results provide valuable insights into the performances of room-temperature macroscopic mechanical systems and their applications in hybrid quantum devices.
\end{abstract}

\maketitle

\section{\label{Introduction}Introduction}

Within the realm of optomechanical coupling \cite{Aspelmeyer_2014}, 
Brillouin scattering in pristine crystals \cite{Brillouin_1922,mandelstam_1926,Chiao_1964} stands out as a particularly promising avenue. 
This mechanism, driven by electrostrictive forces in high-intensity electromagnetic fields \cite{Boyd_2008_2}, enables the transformation of photon energy into mechanical vibrations within the material's lattice. Stokes as well as Anti-Stokes scattering processes are involved in this transformation, and when placed inside an optical cavity, the scattering efficiency is significantly enhanced \cite{Chu_2020}. 

Brillouin scattering has broad implications for quantum storage, quantum sensing, and quantum control. Notably, it offers promising avenues for ground-state cooling \cite{Brubaker_2022,Zhu_2023}, optical-to-microwave transduction \cite{Higginbotham_2018,Zhou2023}, single-phonon level quantum control \cite{Enzian_2021}, and entangling remote mechanical oscillators \cite{riedinger2018remote}, highlighting its significant role in advancing quantum optomechanics \cite{Aspelmeyer_2014}.

Given the potential of harnessing Brillouin interactions for quantum applications, many physical platforms have been explored. These include silica microspheres \cite{Fiore_2011}, silicon photonic waveguides \cite{Eichenfield_2009,Merklein_2017}, and whispering-gallery-mode micro-resonators \cite{Enzian_2018}. 
Among these, bulk acoustic-wave resonators seem particularly promising as they offer several unique advantages, making them highly suitable for practical applications \cite{Aspelmeyer_2014}.
They support ultra-long-lived high-frequency phonon modes with extremely low thermal occupations at helium dilution refrigerator temperatures \cite{Renninger_2018}. Moreover, the gigahertz frequency difference between the pumped and scattered optical modes simplifies mode filtering using conventional optical technologies \cite{Enzian_2021}. Bulk crystals also support multi-wavelength operations, thereby extending their utility for quantum-information processing and wavelength conversion applications \cite{Han_2023}. Finally, as the mechanical modes are massive, these systems serve as an ideal platform for probing macroscopic quantum effects \cite{Thomas_2020}.

Another key benefit of using high-frequency oscillators inside cavities lies in their ability to achieve simultaneous resonance for both the driving and scattered fields \cite{Rueda_2016,Kharel_2019}. This synchronicity significantly boosts the number of photons inside the cavity and, thus, the optomechanical coupling rates. Such a substantial increase is critical for effective phonon control and reaching the strong-coupling regime \cite{Aspelmeyer_2014}. Achieving this regime is vital for demonstrating coherent information transfer between the electromagnetic and the mechanical domains.

Strong coupling has been successfully demonstrated in several optomechanical systems \cite{Groblacher_2009,Aspelmeyer_2014,Sommer_2021,Kharel_2022}, with most implementations utilizing cryogenic-temperature environments. These environments effectively reduce the mechanical linewidth, thereby lowering the threshold for achieving strong coupling. 
Contrarily, in a room temperature environment, the mechanical bandwidth is significantly higher, and, thus, strong coupling can be attained only for very strong interaction strengths. With bulk acoustic modes, this has been demonstrated in a whispering gallery mode resonator but without complete control over the cavity modes \cite{Enzian_2018}. 

In this article, we demonstrate optoacoustic strong coupling in a room temperature environment using a system with full optical control, thus enabling a complete characterization of normal-mode splitting and an avoided crossing. 
Our approach utilizes a pristine quartz crystal that supports high-frequency bulk acoustic phonons at $\Omega_m=\SI{12.43}{GHz}$ and is placed inside a high-finesse Fabry-Pérot cavity to enhance the optomechanical interaction. The setup features a \SIadj{1550}{nm} resonantly-enhanced optical pump 
that drives Anti-Stokes scattering at the signal mode frequency $\omega_{s,0}$ while suppressing the undesired Stokes interaction. Additionally, we employ a Peltier cell to control the cavity temperature, which, in turn, adjusts the cavity length and its detuning from the Brillouin mechanical frequency.  
Our analysis of the optomechanical spectral response involves a heterodyne detection scheme, enabling a comprehensive characterization of the system's performance at room temperature.

\section{Experimental setup and methodology}
A schematic of our fiber and free-space experimental setup is shown in Figure \ref{fig:scheme}. A tunable laser generates the pump mode $\omega_p$ locked to one of the cavity resonances, i.e., $\omega_{p,0}$, with controllable power up to $P_{in}=\SI{300}{mW}$. The signal mode $\omega_s=\omega_p\pm2\pi{f}$ is generated through an EOM modulator with a variable frequency to scan across the higher-frequency cavity resonance $\omega_{s,0}$ and characterize the Anti-Stokes response. The light is coupled to the free-space setup, focused on the \SIadj{4}{mm} z-cut flat-faceted quartz crystal inside the optical cavity. Polarization controllers guarantee the impinging field matches the birefringence axis of the crystal.

The free-space cavity comprises two high-reflectivity mirrors mounted on a brass support that expands with temperature changes (Figure \ref{fig:cavity}). We control the temperature of the cavity using a Peltier element, and, due to the design of the support, this allows for precise tuning of the cavity length $L_c\approx \SI{10.2}{mm}$ while simultaneously reducing the lateral movements of the optical elements and asymmetric pressure forces. It allows for the fine-tuning of the FSR of the selected optical modes to match the Brillouin frequency $\Omega_m$ for in-resonance measurements, and it enables independent adjustments of the detuning $\Delta$ between these modes for the complete response characterization (Figure \ref{fig:modes}).

In our measurement scheme, the light reflected from the cavity (blue line) is coupled back into the optical fiber. Here, it interferes with the local oscillator (red line), i.e., a \SIadj{80}{MHz} shifted replica of the pump signal, and it is then measured with a high-bandwidth (\SI{20}{GHz}) detector. Subsequently, an electronic mixer downconverts the Stokes or Anti-Stokes signal to $\SI{1}{GHz}\pm\SI{80}{MHz}$ before being analyzed using a spectrum analyzer. By varying the sideband modulation frequency $f$, we probe the cavity resonance across the desired frequency range, enabling us to reconstruct the cavity's spectral response.

The reflections from the quartz surfaces result in an asymmetric Free Spectral Range (FSR) spectrum \cite{Kharel_2019}, which allows us to select and resonantly enhance a single scattering process. We focus on the Anti-Stokes response of the system, which is isolated using the beating frequency with the local oscillator.
\noindent
\begin{minipage}{\linewidth}  
\centering
\includegraphics[width=\linewidth]{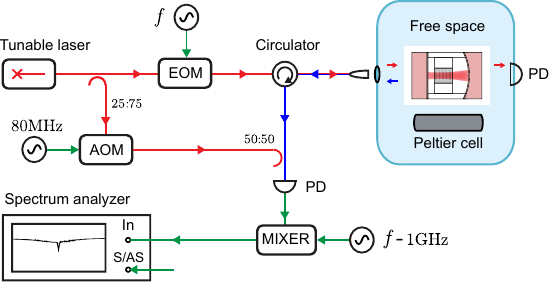}             
\captionof{figure}{\label{fig:scheme}\small Schematic of the measurement setup. A tunable laser generates the pump mode $\omega_p$, locked to a cavity resonance. The signal mode $\omega_s$ is derived from the pumping laser via an Electro-Optic Modulator (EOM) set to a scanning frequency $f$. The light is coupled into the optical cavity and subsequently reflected back into optical fibers, facilitated by a combination of collimators and free-space lenses. The coherent optomechanical response is captured by a heterodyne detector, which beats the reflected probing light against an Acousto-Optic Modulator (AOM)-shifted local oscillator. Finally, an electronic mixer downconverts the signal to $\simeq\SI{1}{GHz}$, enabling a reconstruction of the cavity's spectral response from time-dependent measurements.}
\vspace{\baselineskip}
\includegraphics[width=.6\linewidth]{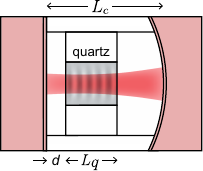}
\captionof{figure}{\label{fig:cavity}\small Schematic of the optomechanical macroscopic cavity configuration, featuring a quartz crystal positioned between two high-reflectivity mirrors and held in place by a brass structure. A Peltier cell mounted atop the cavity regulates temperature changes, allowing for precise calibration and detuning of the optical resonances. The geometrical parameters of the setup include the cavity length $L_c=\SI{10.2}{mm}$, the crystal length $L_q=\SI{4.0}{mm}$, and the distance between the flat mirror and the crystal $d\simeq\SI{1}{mm}$.}
\vspace{.5\baselineskip}
\end{minipage}

Moreover, our system includes a stable and robust locking system that compensates for cavity drifts, ensuring that the laser pump frequency remains in resonance with the cavity mode ($\omega_p=\omega_{p,0}$). To minimize measurement noise and accurately quantify pump power fluctuations, we analyze the lock traces and fine-tune the lock's PID parameters before starting each data acquisition.
Fluctuations in the power coupled to the cavity, particularly while varying the temperature, can impede accurate estimation of the optomechanical parameters. This is due to the measurement bias introduced when collecting data at different detunings. 
Additionally, fluctuations become increasingly problematic at higher laser powers, where a small relative change in power corresponds to a large absolute change that the temperature control must offset, and limits our maximum input power to about \SI{300}{mW}.

\section{Theoretical model for the cavity reflectivity}

In the description of our experiment, we consider the evolution of two optical modes cavity modes described by the creation and annihilation operators $\hat{a}_p$ and $\hat{a}_s$, with frequencies $\omega_p$ and $\omega_s$, which are coupled through a mechanical mode $\hat{b}$ of frequency $\Omega_m$ \cite{Aspelmeyer_2014}. The pump mode is driven by a strong laser field at frequency $\omega_p$, while a weak probe field is used to scan through the mode $\hat{a}_s$ and monitor its response. The system is additionally characterized by the decay rates $\kappa_p$, $\kappa_s$, and $\Gamma_m$, respectively corresponding to the pump, signal, and mechanical modes. 

The starting point for our description is the Hamiltonian 
\begin{equation}
    \hat{H} = \hat{H}_{0} + \hat{H}_{l_d} + \hat{H}_{l_p} + \hat{H}_{int}
\label{eqn:hamiltonian_short}
\end{equation}
where the bare energy of the system corresponds to $\hat{H}_{0} = \hbar \omega_p \hat{a}_p^\dagger\hat{a}_p + \hbar \omega_s \hat{a}_s^\dagger\hat{a}_s + \hbar \Omega_m \hat{b}_m^\dagger\hat{b}_m$, the laser driving terms are $\hat{H}_{l_i} = i \hbar \alpha_i ( \hat{a}_i^\dagger e^{-i \omega_i t} - \hat{a}_i e^{i \omega_i t} )$ for $i = p, s$, and $\hat{H}_{int}$ corresponds to the optomechanical interaction. 
In general, the optomechanical interaction term $\hat{H}_{int}$ would contain both the Stokes and anti-Stokes interactions for each of the cavity fields, but note that only the anti-Stokes of the driving mode will be in resonance and that the probe field is weak compared to the drive field we can simplify it to
\begin{equation}
    \hat{H}_{int} = -\hbar g_0 \hat{a}_s^\dagger \hat{a}_p \hat{b}_m + \text{h.c.}
\label{eqn:hamiltonian_int}
\end{equation}
The strength of the pump field leads to a high intracavity power, which lets us treat the mode  $\hat{a}_p$ classically and use the undepleted pump approximation \cite{Kharel_2019}.
Following a conventional framework for coupled optomechanical systems \cite{Kharel_2019}, we use a rotating reference frame ($\hat{a}_s(t) \rightarrow \hat{a}_s(t) e^{-i \omega_p t}$ and $\hat{b}_m(t) \rightarrow \hat{b}_m(t) e^{-i (\omega_p - \omega_s) t}$, and upon adopting the rotating frame approximation the effective linearized Hamiltonian can be written as
\begin{equation}
\begin{split}
    \hat{H}&=-\hbar(\Omega-\Omega_m)\hat{b}_m^\dagger\hat{b}_m- \hbar(\Omega-\Omega_m-\Delta)\hat{a}_s^\dagger\hat{a}_s\\
    &-\hbar g_m(\hat{a}_s^\dagger\hat{b}_m+\hat{a}_s\hat{b}_m^\dagger)
    +i\hbar\sqrt{\kappa_{ext,s}}\alpha_{in,s}(\hat{a}_s^\dagger-\hat{a}_s)\\
\end{split}
\label{eqn:hamiltonian}
\end{equation}

If the effective coupling strength $g_m$ does not affect the mode dynamics, the pump mode remains undepleted and the coupling rate is proportional to the amplitude of the intracavity photon field and the impinging pump field, expressed as $g_m=g_0{N}^{1/2}\propto{P_{in}^{1/2}}$ where $g_0$ represents the single-photon coupling rate. 
The final term in the system Hamiltonian describes the impinging field with amplitude $\alpha_{in,s}$, which facilitates coupling power into the cavity through the first mirror ($\kappa_{ext,s}$).

\noindent
\begin{minipage}{\linewidth}  
\centering
\includegraphics[width=.8\linewidth]{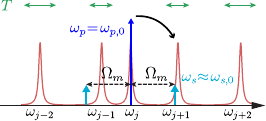}
\captionof{figure}{\label{fig:modes}\small  This pictogram represents the distribution of asymmetric cavity modes when the pump laser frequency is locked to a specific cavity mode $\omega_p=\omega_{p,0}=\omega_j$. By adjusting the cavity's FSR to align the higher frequency mode to match the Brillouin frequency $\Omega_m$, we facilitate the generation of signal sidebands near $\Omega\simeq\Omega_m$. This alignment allows for the scanning over the cavity mode and the activation of the Anti-Stokes optomechanical interaction. Notably, the lower frequency sideband is suppressed, as it does not resonate with any cavity mode.
}
\vspace{.5\baselineskip}
\end{minipage}

We derive the Langevin equations of motion for the optical and acoustic fields. The steady-state solution for the intracavity fields is given by 
\begin{equation}
\begin{cases}
\displaystyle
    \hat{b}_m=\frac{ig_m}{\frac{\Gamma_m}{2}-i(\Omega-\Omega_m)}\hat{a}_s\\
\displaystyle
    \hat{a}_s=\frac{\sqrt{\kappa_{ext,s}}\alpha_{in,s}}{\frac{\kappa_s}{2}-i(\Omega-\Omega_m-\Delta)-\frac{g_m^2}{\Gamma_m/2-i(\Omega-\Omega_m)}}
\end{cases}
    \label{eq: complete steady state}
\end{equation}

To further develop our model, we consider the cavity reflectivity $R(\Delta,\Omega)$. This is done by considering the input-output relation between the input and output cavity modes $\hat{a}_{in,s}$ and $\hat{a}_{out,s}$, \(\hat{a}_{out,s}=\hat{a}_{in,s}-\sqrt{\kappa_{ext,s}}\hat{a}_s\).  We also incorporate the scattering matrix \(S\), which links the free-space fields with those in the cavity, as described in \cite{Doeleman_2023}. The resulting expression for the optomechanically induced transmissivity (OMIT) depends on the optical and mechanical linewidths as well as the coupling strength $g_m$.
\begin{equation}
\begin{split}
    R(\Delta,\Omega)=& \abs{S_{12}S_{21}\frac{\braket{\hat{a}_{out,s}(\Delta,\Omega)}}{\braket{\hat{a}_{in,s}(\Delta,\Omega)}}+ S_{11}}^2\\
    =&\abs{1-\frac{A\kappa_{s}/2}{\frac{\kappa_s}{2}-i(\Omega-\Omega_m-\Delta)-\frac{g_m^2}{\Gamma_m/2-i(\Omega-\Omega_m)}}}^2
\end{split}
\label{eq: model}
\end{equation}
Note that $g_m$ is the only parameter in this model dependent on power. In our final expression, we normalize the reflected signal to the impinging mode. As a result, the factor $A<1$ only depends on the optical linewidth $\kappa_s$, the coupling coefficient $\kappa_{ext,s}$, and the cavity mode matching described by the matrix $S$. Moreover, by spectral characterizing the impinging power, we effectively reduce the impact of external low-finesse cavities that may arise from multiple reflections from the free-space optical components.

\noindent
\begin{minipage}{\linewidth}  
\centering
\includegraphics[width=.98\linewidth]{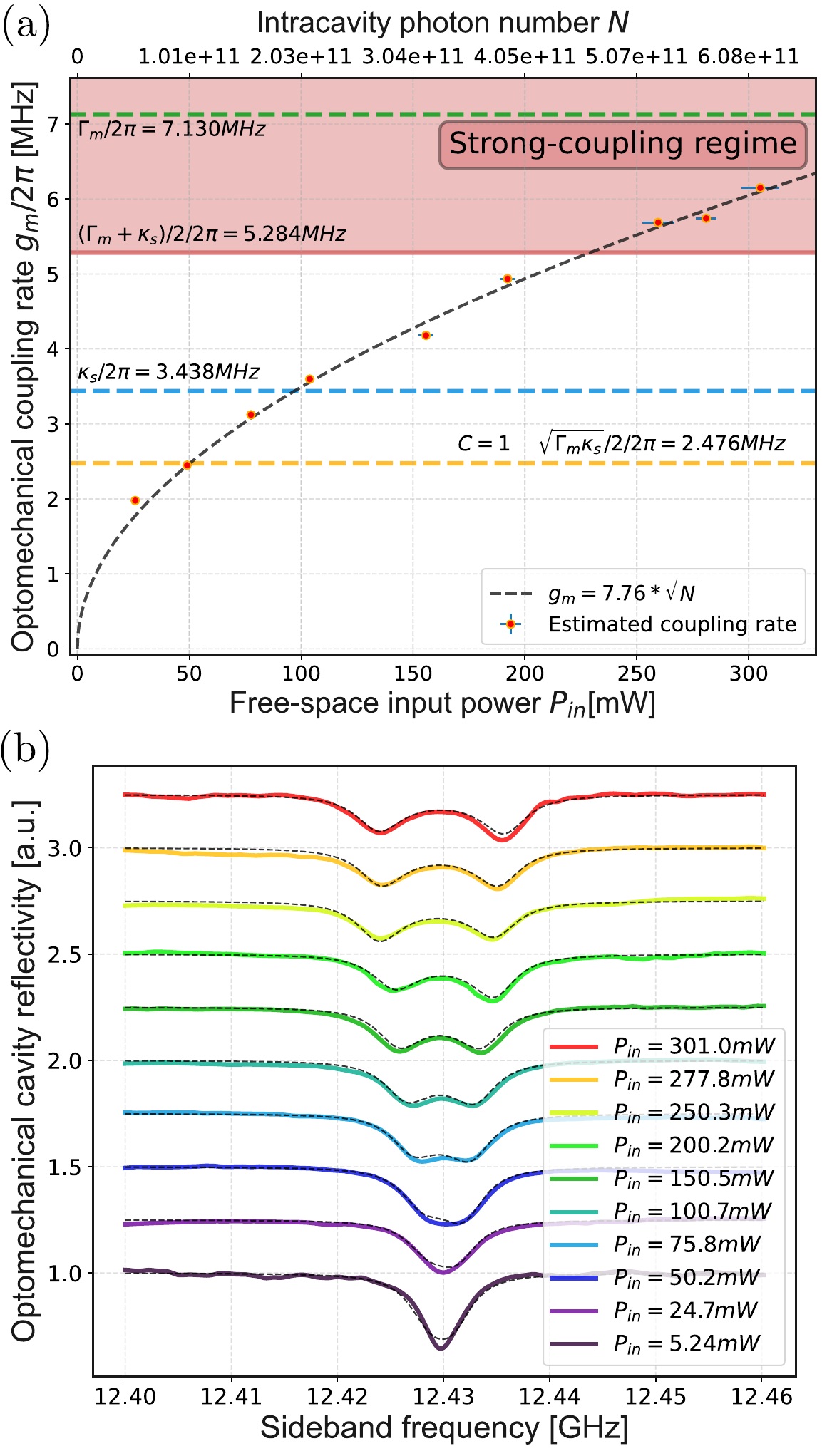}
\captionof{figure}{\label{fig:strongcoupling}\label{fig:inresonance}\small (a) Power-dependent spectral response of the optomechanical cavity. As the power input increases, the cavity's resonance peak undergoes a splitting and separation, scaling with the square root of the input power $P_{in}^{1/2}$. The system transitions into the strong-coupling regime when the optomechanical coupling rate $g_m$ surpasses the average of the optical $\kappa_s$ and mechanical $\Gamma_m$ linewidths. The transition is when $P_{in}$ exceeds $\SI{239}{mW}$ and leads to a resolution of the two peaks. 
(b) Plotting the derived single-photon coupling rate as a function of input power up to \SI{300}{mW} clearly demonstrates the optomechanical coupling strength's square-root scaling with the intracavity photon number. From our data analysis, we derive a single-photon coupling rate $g_0$ of $\SIpm{7.76}{0.05}{Hz}$. Notably, the effective coupling rate $g_m$ reaches the threshold for the strong-coupling regime when the input power $P_{in}$ exceeds $\SI{239}{mW}$, as highlighted in the shaded pink region of the graph.
}
\vspace{.5\baselineskip}
\end{minipage}
\noindent
\begin{minipage}{\linewidth}
    \centering
    \captionof{table}{\small Cavity parameters achieving strong coupling.}
    \label{tab: results}
    \centering
    \scalebox{1}{
        \begin{tabular}{ll}
        \hline  
        $\mathrm{Parameter}$ & $\mathrm{Value}$ \\
        \hline
        $\omega_p/2\pi\approx\omega_s/2\pi$ & \SI{194.3}{THz}\\
        $\lambda_p\approx\lambda_s$ & \SI{1543}{nm}\\
        $\mathrm{Cavity~temperature}~T$ & \SIrange{22.7}{24.3}{^\circ{C}}\\
        $\Omega_m/2\pi$&\SI{12.43}{GHz}\\
        $\kappa_p/2\pi$&\SI{3.029}{MHz}\\
        $\kappa_s/2\pi$&\SI{3.438}{MHz}\\
        $\Gamma_m/2\pi$&\SI{7.130}{MHz}\\
        $\mathrm{Cavity~Finesse}~F$ & $\gtrsim\SI{3500}{}$\\
        $\mathrm{Pump~quality~factor}~Q_p$ & \SI{6.3e8}{}\\
        $\mathrm{Signal~quality~factor}~Q_s$ & \SI{5.7e8}{}\\
        $g_0/2\pi$ & \SIpm{7.76}{0.05}{Hz}\\
        $\max \{g_m/2\pi\}$ & \SI{6.15}{MHz}\\
        $\max \{P_{in}\}$ & \SI{301}{mW}\\
        $\max \{N\}$ & \SI{6.1e11}{}\\
        \hline
        \end{tabular}
    }
\end{minipage}

\section{Analysis of strong coupling}
Strong coupling refers to a distinct regime where the interaction between the optical field and the mechanical vibrations enables an efficient and coherent energy exchange. This interaction results in the formation of collected excited states known as photon-phonon hybrid states \cite{Aspelmeyer_2014}. To achieve strong coupling, the optomechanical coupling rate $g_m$ must exceed the sum of the intrinsic decay rates of both the optical and mechanical modes, i.e., $g_m>(\kappa_s+\Gamma_m)/2$.

Our experimental setup allows for the independent adjustment of the detuning between the cavity mode FSR and Brillouin frequency, as well as the pumping power. This capability enables us to investigate the impact of the effect of the coupling strength on the in-resonance ($\mathrm{FSR} = \Omega_m$) spectral measurements, as shown in Figure \ref{fig:inresonance}. In the weak-coupling regime $g_m\ll \kappa_s$, the optomechanical interaction primarily causes an increase in the effective linewidth of the cavity mode. However, as the coupling rate surpasses $\abs{\kappa_s-\Gamma_m}/4$, the initial Lorentzian optical response evolves into a normal-mode splitting of the transmission peak. With increasing intracavity photon number, the peak separation approaches $2g_m\sim N^{1/2}$, indicating that the system's response is not only governed by the optical system but also by the entire optomechanical interaction. In this scenario, the system's eigenmodes correspond to the new annihilation operators $(\hat{a}_s \pm \hat{b})/\sqrt{2}$. The transition to the strong-coupling regime is marked when the peaks are resolved, i.e., when $g_m>(\kappa_s+\Gamma_m)/2$.

In our analysis of the optomechanical system, we go beyond the characterization of in-resonance spectra to include measurements under varying FSR detunings and pumping powers. This approach enables us to estimate the Brillouin frequency and the mechanical linewidth. 
Additionally, we can determine the single photon coupling rate by examining the relationship between the coupling strength and the intracavity photon 
\noindent
\begin{minipage}{\linewidth}
\centering
\includegraphics[width=.95\linewidth]{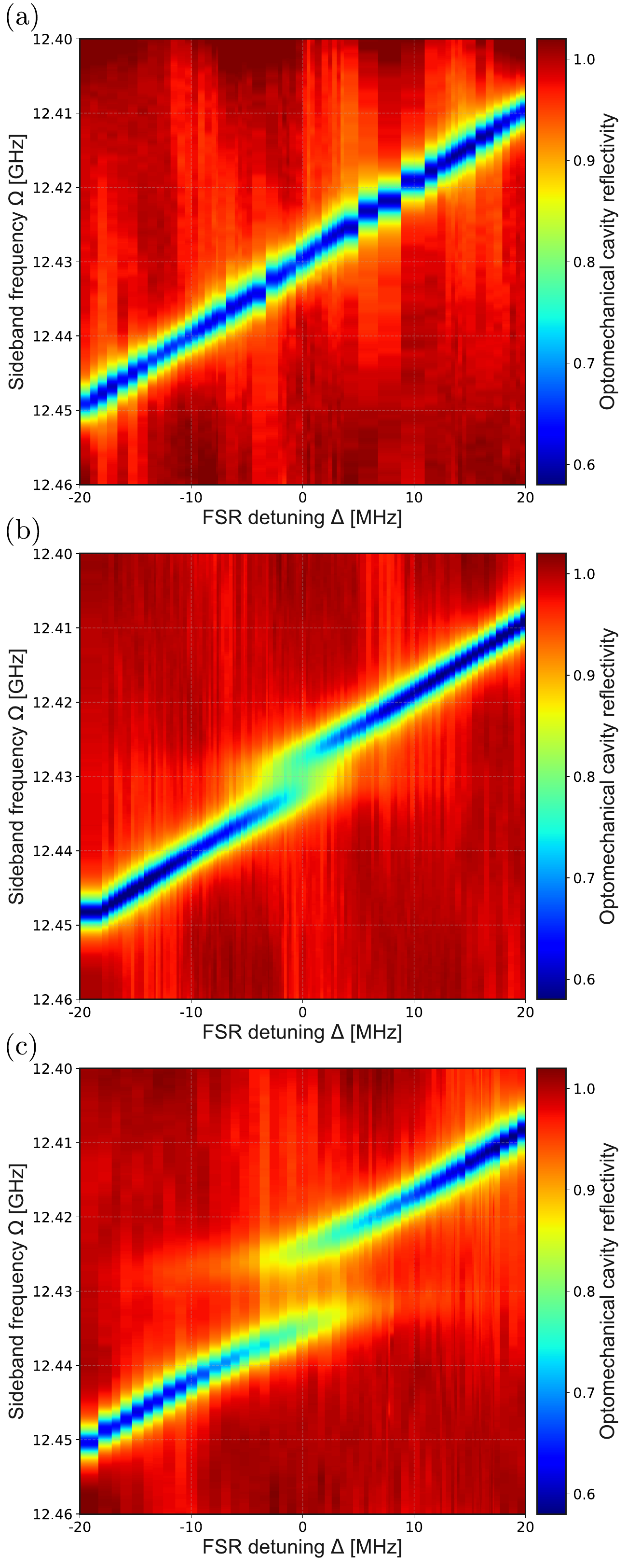}
\captionof{figure}{\small \label{fig:detuning}Spectra of the cavity's optomechanical response for varying pumping powers $P_{in}~=~\SI{5.24}{mW},$ $\SI{75.8}{mW},$ $\mathrm{and}$ $\SI{277.8}{mW}$. Each plot shows the spectral response as a function of the FSR detuning $\Delta$. With an increasing optomechanical coupling rate, the spectral response evolves from a single peak into a dual-peak structure. This transition culminates in the strong-coupling regime, where the plots distinctly exhibit an avoided crossing.}
\end{minipage}
number. 
We simultaneously fit all datasets to the same theoretical model, allowing only the coupling strength to change for different input powers, and the detuning to change within the same input power.
The derived parameters are presented in Table \ref{tab: results} along with the key cavity attributes. Notably, the measured Brillouin frequency $\Omega_m/2\pi=\SI{12.43}{GHz}$ matches the expectation for our quartz crystal \cite{Renninger_2018}, confirming its independence from both the pumping power and the optomechanical coupling rate.
It is significant to mention that the optical linewidth $\kappa_s$ can also be estimated using the same optimization algorithm but with out-off-resonance spectra. These spectra are not influenced by the mechanical linewidth or the coupling strength, providing a distinct advantage. While the linewidths of both pump and signal modes can also be verified by directly detecting the cavity reflectivity without locking the laser, this first approach offers higher accuracy. This is because the measurement conditions - specifically, the cavity temperature and thermal distribution - are consistent with those used in the experimental data for the optomechanical response.

The optomechanical coupling rate estimated at the different power levels is displayed in Figure \ref{fig:strongcoupling}. This graph validates the square-root relationship between the pump laser power and the Brillouin interaction, enabling us to estimate the single-photon coupling rate as $g_0/2\pi=\SIpm{7.76}{0.05}{Hz}$. This value falls within the theoretical prediction of $g_0/2\pi\lesssim\SI{13}{Hz}$, which is achievable with perfect mode overlap and phase matching \cite{Kharel_2019}. The power impinging on the cavity and the number of intracavity photons for each data point were deduced from the reflected and transmitted power measurements taken before and during the experimental process. This strategy also confirms the stability of the power during data acquisition and allows for the estimation of fluctuations attributable to cavity mode instabilities, particularly noticeable above \SI{150}{mW}. The figure also indicates the strong-coupling threshold, calculated to be \SI{5.284}{MHz}, demonstrating the strong-coupling regime achieved for $P_{in}>\SI{239}{mW}$.

Another signature of strong coupling is the spectra evolution with the FSR detuning. In Figure \ref{fig:detuning}, we present 2D maps of the cavity reflectivity, plotted as a function of the sideband scanning frequency $\Omega$ and the FSR deviation from the Brillouin frequency $\Delta$. At low powers, such as $P_{in}=\SI{5.24}{mW}$ (Figure \ref{fig:detuning}a), a change in temperature increases the cavity FSR, shifting the central frequency of the Lorentzian transmission peak across the scanning range. As the pump power increases, e.g., $P_{in}=\SI{75.8}{mW}$ in Figure \ref{fig:detuning}b, we observe a normal-mode splitting in the peak, a separation that increases proportionally to $P_{in}^{1/2}$. Finally, in Figure \ref{fig:detuning}c, at $P_{in}=\SI{277.8}{mW}$, the system enters the strong-coupling regime, characterized by the distinct avoided-crossing pattern. Here, in resonance, the peaks are resolved because the splitting exceeds twice the hybrid-mode linewidth. This prevents the peaks from crossing at the FSR detuning, resulting in the observed avoided-crossing.
\section{Conclusions}
In this study, we have successfully demonstrated and characterized the room-temperature strong-coupling regime in a \SIadj{4}{mm} quartz crystal, facilitated by Brillouin optomechanical interaction with high-frequency (\SI{12.43}{GHz}) phonons. Enclosing the crystal into a free-space optical cavity enables the probing of multiple resonance modes. It allows for in-resonance optical pumping, thus reducing the power required to achieve strong coupling. Despite a modest single-photon coupling rate of \SI{7.76}{Hz} and the limitations of a less efficient acoustic cavity ($\Gamma_m/2\pi=\SI{7.13}{MHz}$), we achieved an optomechanical coupling rate in the \SI{}{MHz}-range. This accomplishment surpasses the strong-coupling threshold for an impinging power exceeding $P_{in}>\SI{239}{mW}$. Our systematic characterization permits the independent control of the coupling strength and the FSR detuning, clearly revealing normal-mode splitting and an avoided crossing in the optical reflectivity spectrum -- definitive indicators of operation within the strong-coupling regime.

Future improvements to our setup include exploring crystals with higher Brillouin gains, such as $\mathrm{TeO_2}$ \cite{Renninger_2018}, which may facilitate access to the strong-coupling regime. Additionally, improving the stability at high powers through enhanced thermalization of the optical cavity and a faster locking system \cite{Black_2001} is also a potential area of further development. Beyond these technical improvements, our experimental setup paves the way for studies in the quantum-coherent strong-coupling regime \cite{verhagen_2012}. By transitioning the setup to cryogenic temperatures, we aim to reduce the mechanical damping rate, extend the phonon lifetime, and reach the mechanical ground state. The low-temperature environment would allow the optomechanical coupling rate to surpass thermal decoherence ($g_m>{\Gamma_m N_{th}}$, where $N_{th}$ is the average number of thermal photons), thus enabling coherent exchange of quantum states between the optical and acoustic modes \cite{Aspelmeyer_2014,Higginbotham_2018}. Achieving this would be a pivotal step towards the development of quantum memories and photon-phonon interfaces in the \SIrange{5}{100}{GHz} range, as well as other Brillouin-based hybrid quantum devices.

\section*{Acknowledgements}
This work has received funding from the Danish National Research Foundation (bigQ DNRF 142).
The authors declare that they have no competing interests.
All data needed to evaluate the conclusions in this paper are presented in the paper. 
Additional data are available from the authors upon request.

\bibliographystyle{apsrev4-2}
\bibliography{bibliography}

\end{document}